# Weighted Least Squares (WLS) Density Integration for Background Oriented Schlieren (BOS)

**Lalit Rajendran[1§], Jiacheng Zhang[2§], Sally Bane[1], Pavlos Vlachos[2*]**

[1] Purdue University, School of Aeronautics and Astronautics, West Lafayette, USA.

[2] Purdue University, School of Mechanical Engineering, West Lafayette, USA.

*pvlachos@purdue.edu

**Abstract**

We propose an improved density integration methodology for Background Oriented Schlieren (BOS) measurements that overcomes the noise sensitivity of the commonly used Poisson solver. The method employs a weighted least-squares (WLS) optimization of the 2D integration of the density gradient field by solving an over-determined system of equations. Weights are assigned to the grid points based on density gradient uncertainties to ensure that a less reliable measurement point has less effect on the integration procedure. Synthetic image analysis with a Gaussian density field shows that WLS constrains the propagation of random error and reduces it by 80% in comparison to Poisson for the highest noise level. Using WLS with experimental BOS measurements of flow induced by a spark plasma discharge show a 30% reduction in density uncertainty in comparison to Poisson, thereby increasing the overall precision of the BOS density measurements.

[§] These authors contributed equally to this work.

# 1. Introduction and Methodology

Background Oriented Schlieren (BOS) is an optical technique used to measure density gradients by tracking the apparent distortion of a target dot pattern [1]. The apparent displacement is obtained by comparing the distorted image and a reference image without the density gradients, and the estimation can be performed by cross-correlation, tracking, or optical flow algorithms [1]–[3]. This displacement is related to the density gradient field and the optical layout parameters as given by

$$\Delta \vec{X} = \frac{MZ_D K}{n_0} \int \nabla \rho \, dz \qquad (1)$$

where $\Delta \vec{X}$ is the apparent displacement, $M$ is the magnification of the dot pattern, $Z_D$ is the distance between the dot pattern and the mid plane of the density gradient field, $K$ is the Gladstone-Dale constant (= $0.225 \times 10^{-3}$ kg/m³ for air), $n_0$ is the ambient refractive index, $\nabla \rho$ is the density gradient field and $z$ is the co-ordinate along the viewing axis. The integral is over the depth/thickness of the density gradient field.

Given the apparent displacement from the image processing algorithms, Equation (1) can be used to calculate the projected density gradient field, and then the density field can be obtained by spatial integration [4]. The BOS experimental setup is simple, easy to use, yields quantitative density information as opposed to the traditional schlieren technique [5] and can be extended to large scale flows.

The density gradient integration is traditionally performed by solving Equation (2) using a Poisson solver,

$$\begin{aligned} \rho &= (\nabla^2)^{-1}(\nabla \cdot \nabla \rho) \\ &= (G^T G)^{-1}(G^T \nabla \rho) \end{aligned} \qquad (2)$$

where $\rho$ is the density field, $\nabla \rho$ is the density gradient field, and $G$ is the gradient operator used for discretizing the derivative [4]. However this procedure is sensitive to measurement noise, and the noise can spread from one part of the measurement domain to contaminate other regions [6]. There can be several sources of noise in BOS measurements. For example, non-uniform illumination in the field of view that can increase the effect of image noise, unreliable measurements due to a failure in the displacement estimation algorithm, as well as noise in the boundary condition such as to a pressure/temperature measurement from a probe. Therefore, a robust integration method is required that can withstand and constrain the effect and propagation of measurement noise.

Least Squares (LS) Optimization is an alternate approach that is robust and customizable. It involves formulating the integration as an optimization problem, with the aim of minimizing a pre-defined cost function. For example, the cost function can be defined as the difference between the measured density gradient field and a finite difference approximation of the unknown density field, and the density field can be calculated by minimizing this cost function subject to constraints imposed by the boundary conditions. While solving the least squares problem with the particular cost function defined above is mathematically equivalent to solving the Poisson equation in (2) given the same stencil, the advantage of the LS-optimization based approach is that it allows the introduction of additional information/constraints about the flow field and flow measurement to improve the density integration procedure.

For example, non-uniform weights can be assigned to grid points to form the Weighted Least Squares (WLS) problem, which can be solved by Equation (3), where $W$ is the "weight matrix".

$$\rho = (G^T W G)^{-1}(G^T W \nabla \rho) \quad . \quad (3)$$

A common approach is to assign weights based on the inverse variance of the measurement error for each point, to ensure that more precise measurements have a greater effect on the result. Recently Zhang et. al. [7],[8] showed that WLS can significantly improve the performance of velocity-based pressure integration in incompressible flow, when the weights are assigned based on the accuracy of pressure gradient estimated from velocity error or velocity uncertainty. However, this approach is not applicable to planar BOS in general, due to the compressibility of the flow. Instead, the weights can be assigned based on the uncertainty of the BOS measurement which is directly related to the density gradient.

Recently, Rajendran et. al. [9], [10] made significant advancements in uncertainty quantification methods for BOS measurements by developing a method to report local, instantaneous, a-posteriori density uncertainty across all points in the field of view. There has been a recent development in uncertainty quantification method for BOS measurements by Rajendran et. al. [9], [10] to report local, instantaneous, a-posteriori density uncertainty across all points in the field of view. To achieve this, PIV-based displacement uncertainty quantification methods are used to estimate displacement uncertainties from cross-correlation BOS and propagated through the density integration procedure. One of the findings is that displacement uncertainty schemes from PIV are also applicable for cross-correlation BOS, and that result will be used here. Further, the methodology to estimate the density uncertainty will also be utilized in this work.

Therefore, we propose a WLS-based density integration methodology for BOS wherein the displacement/density gradient uncertainty will be used to assign weights for the integration procedure. For a grid point $k$, the weight is given by,

$$\begin{aligned} W_k &= \left(\sigma_{\nabla\rho,k}\right)^{-2} \\ &= \left(\frac{1}{MKZ_D\Delta z}\sigma_{\Delta X,k}\right)^{-2} \end{aligned} \quad , \quad (4)$$

where $\sigma_{\nabla\rho,k}$ is the density gradient uncertainty at this point and $\sigma_{\Delta X,k}$ is the displacement uncertainty. The pre-multiplying term involves the optical layout parameters described earlier in Equation (1), with the additional parameter $\Delta z$ denoting the depth/thickness of the density gradient field. This weight matrix is used along with Equation (3) to perform the WLS density integration for BOS. In this manner, the unreliable density gradient data points (with greater uncertainty) are assigned lower weights as defined in Equation (4), and thus have a lower effect on the density integration procedure. Since the displacement uncertainty calculation procedures are sensitive to a wide variety of error and uncertainty sources such as low illumination, image noise, displacement gradients etc. [11], we are able to identify unreliable measurements in a robust manner. If the errors in the density gradient are unbiased and uncorrelated, the weight matrix is the inverse of the covariance matrix of the density gradient error, and WLS provides the best unbiased linear estimator for the density integration problem [12].

The following sections detail the assessment of this methodology with synthetic and experimental BOS images, and show that WLS can reduce the density random error/uncertainty and improve the overall precision of the measurement.

## 2. Analysis with synthetic BOS images

We performed error analysis using synthetic BOS images with a known density field, and assessed the performance of three density integration algorithms: (1) Poisson solver, (2) WLS with weights based on the random displacement error, and (3) WLS with weights based on the displacement uncertainty. Further, a patch of high noise was added to one part of the BOS image to assess how the error due this patch propagates to the surrounding field during the density integration procedure.

The synthetic BOS images were generated using a ray tracing-based image generation methodology [13]. In this method, light rays are launched from source points on the BOS target, propagated through density gradients using methods taken from the gradient index optics literature, and then through the camera lens up to the final intersection with the camera sensor to generate the dot pattern images.

A Gaussian density field with a peak density offset of 0.3 kg/m$^3$ and a spatial standard deviation of ¼ of the field of view was chosen to render the images, as shown in Figure 1(a). The synthetic images featured a random dot pattern with a dot size of 3 pixels and a dot density of 15 dots/32x32 pixel window, and the entire image is corrupted with a noise level that was 1% of the peak image intensity. In addition, a portion of the image was corrupted with image noise higher than the surrounding regions by a specified amplification ratio. Three amplification ratios are considered in this analysis: 1, 10, 20. In all cases, the image noise at a given pixel was drawn randomly from a Gaussian distribution with the standard deviation determined from the noise level. A sample image highlighting the high-noise region is shown in Figure 1(b). A total of 1000 such image pairs were generated, with each image pair consisting of one image rendered with the density gradient field and a reference image without the field with the same noise level and noise pattern as the gradient image.

Each image pair is cross-correlated using a multi-pass window deformation procedure with 32x32 pixel windows and 50% overlap to obtain the corresponding displacement field shown in Figure 1(c). Then the displacement error is calculated from the deviation of the measured displacements from cross-correlation with respect to the light ray deflections from ray tracing. Finally, the displacement errors from all image pairs are used to compute the error statistics, and the spatial distribution of the random component of the displacement error is shown in Figure 1(e). It is seen to be higher in the noisy patch, as expected.

In addition, the displacement uncertainty is also estimated during the correlation processing using the Moment of Correlation (MC) algorithm developed by Bhattacharya et. al. [14] which estimates the uncertainty from the PDF of displacements that contribute to the cross-correlation plane. The instantaneous displacement uncertainty field is shown in Figure 1(d), and it also shows a large increase in the noisy patch, which is consistent with the increase in the random error. However, the displacement uncertainty underpredicts the random error in the noisy patch and over-predicts the random error in the rest of the field. A point to note is that the random error is a statistically averaged estimate while the MC uncertainty is from a single snapshot. Thus, the weights chosen based on the random error/uncertainties were lower in the noisy patch, and thus measurements in the patch would have less effect on the density integration.

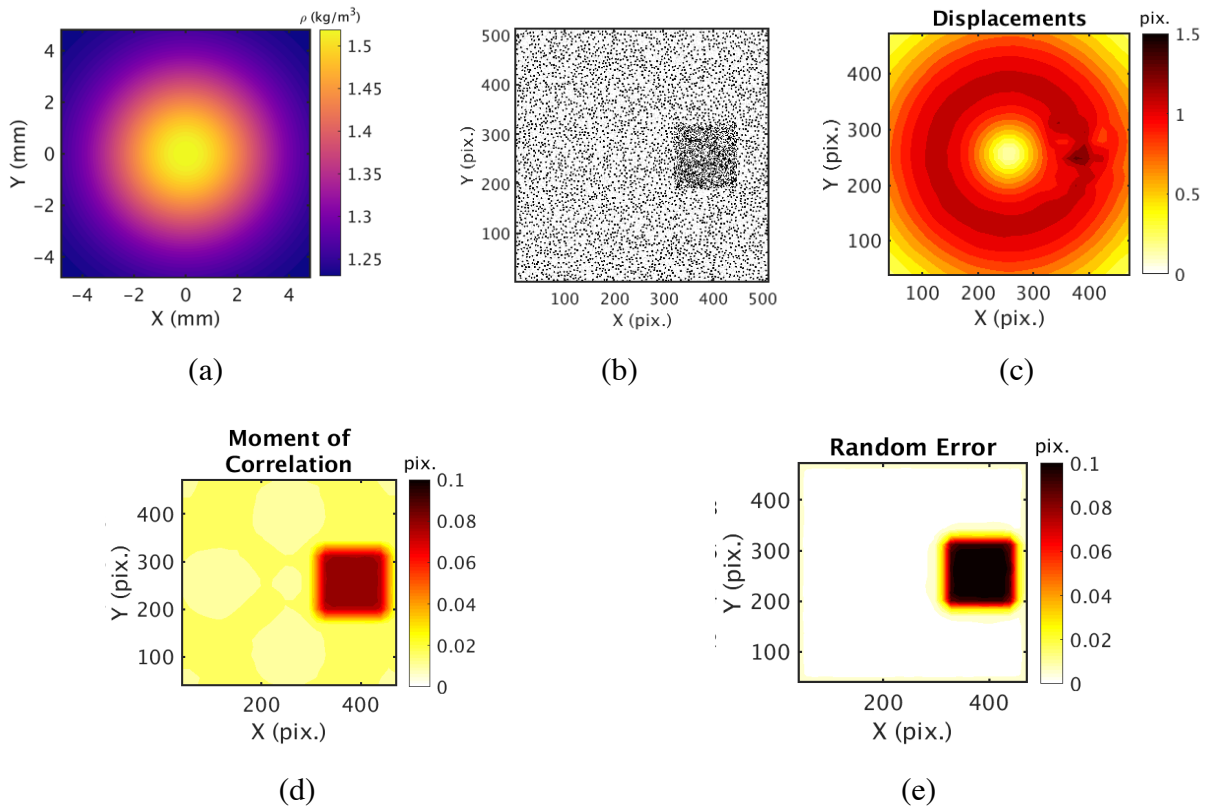
5

Figure 1. Synthetic dataset used for comparing the density integration methods corresponding to an amplification ratio of 20. (a) Gaussian density field, (b) BOS image in false color showing the region of the image corrupted with noise, (c) corresponding displacement field, (d) instantaneous displacement uncertainty from MC, and (e) random error of the displacement from 1000 realizations.

Next, the displacements were used to calculate the density gradients, which were then spatially integrated to calculate the density field using the three integration methods described earlier. A $2^{nd}$ order central difference scheme is used for spatial discretization, with Neumann boundary conditions on all four boundaries and the Dirichlet boundary condition at the midpoint of the top boundary. The density error was then calculated by comparing the integrated density field with the reference field used to render the images, and the error statistics from 1000 such vector fields are calculated.

The resulting random error in the density field obtained from the three integration methods are shown in Figure 2. The WLS method (middle and right columns) is able to constrain the spread of the random error from the patch, while the error has a wider spread with Poisson integration (left column), and this difference increases with the patch amplification ratio. Finally, the results show that WLS with weights based on the displacement uncertainty (middle column) performs just as well as the case with weights based on the random error (right column), thereby justifying the use of the displacement/density gradient uncertainty to estimate the weights. This demonstrates the practical value of WLS since the error is not known and only the uncertainty can be estimated in the vast majority of experiments.

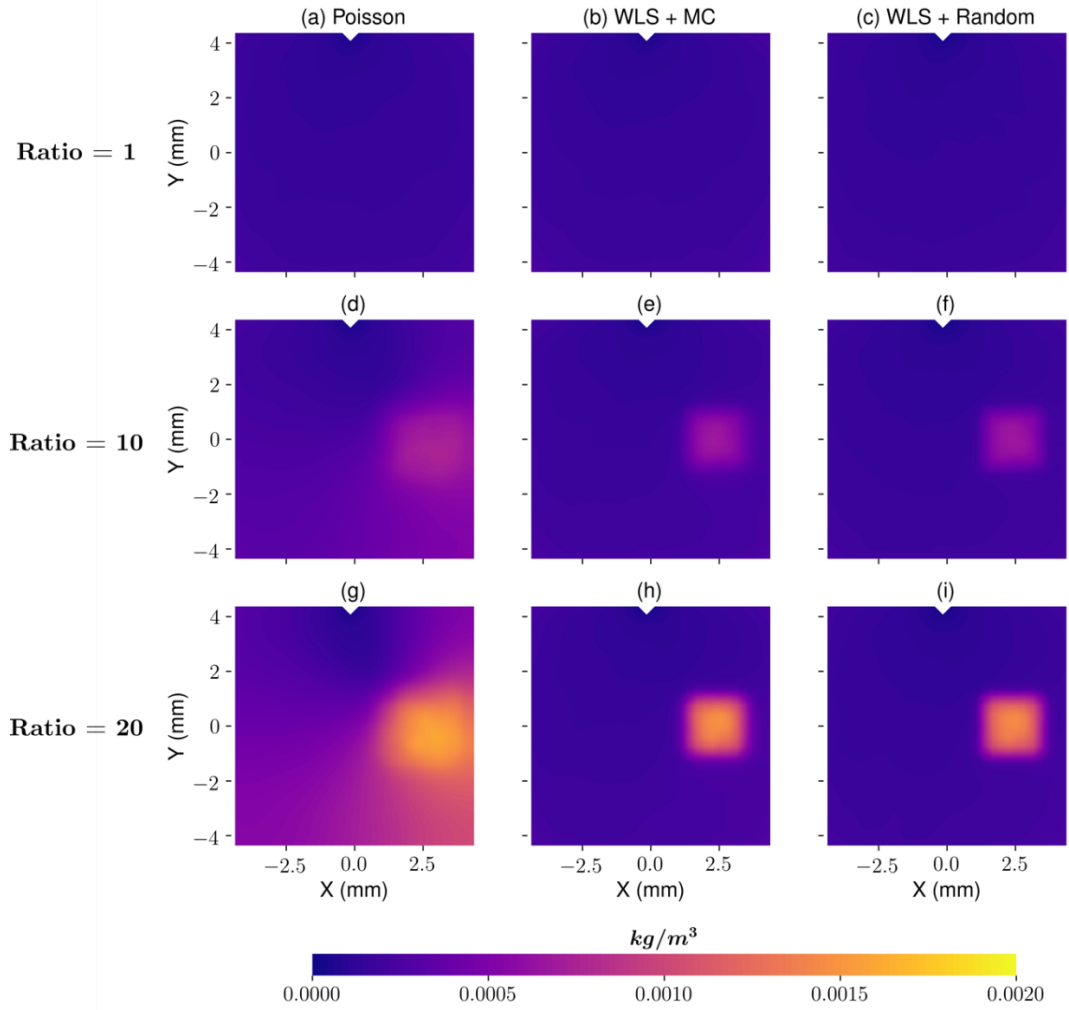

Figure 2. Spatial distribution of the density random error associated with the three integration methods. Each row corresponds to the same patch noise amplification ratio, denoted on the left, and each column corresponds to an integration scheme. (a), (d), (g): Poisson; (b), (e), (h): WLS with weights based on displacement uncertainty from MC; and (c), (f), (i): WLS with weights based on displacement random error.

In addition, the probability density function (PDF) and cumulative density function (CDF) of the density random error distribution was calculated from 250,000 grid points, and are shown in **Error! Reference source not found.**. In addition, the RMS error is shown for the PDF plot and the 90[th] percentile error is shown for the CDF plot. From these distributions, it is seen that as the patch noise amplification ratio increases, WLS results in a greater reduction in both the RMS and 90[th] percentile errors in comparison to Poisson, with a 50% reduction in the RMS value and an 80% reduction in the 90[th] percentile of the density random error for the highest patch amplification ratio. It is also seen that the 90[th] percentile result from WLS is nearly independent of the noise level, which is a direct result of the previous observation that it reduces the spread of random error.

And it is again seen that WLS with weights based on MC uncertainty, results in an error distribution that is practically identical to WLS with weights based on the random error.

In summary, the error analysis shows that the WLS integration can significantly improve the precision of the density integration procedure in comparison to the traditional Poisson solver.

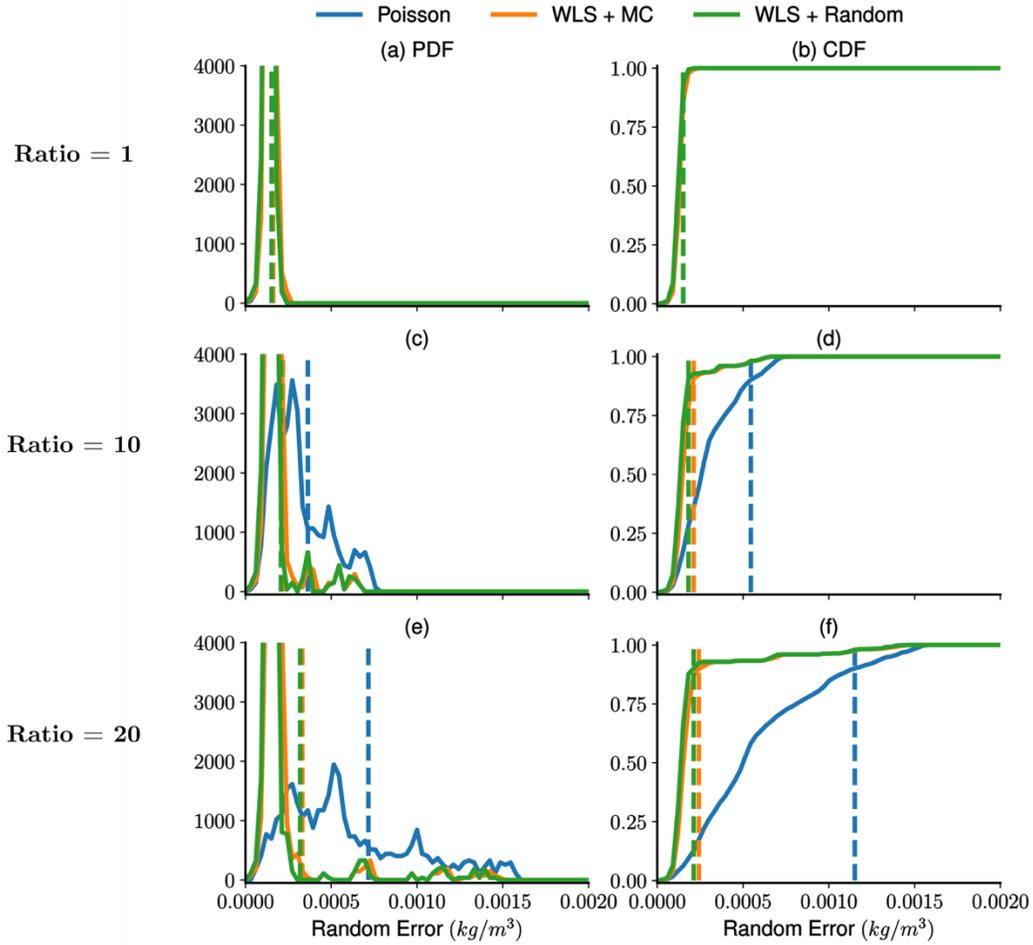

Figure 3. PDF (left) and CDF (right) of the density random error associated with the three integration schemes. The dashed lines indicate the RMS error for the left column, and the 90$^{th}$ error percentile for the right column. Each row corresponds to the same patch noise amplification ratio, denoted on the left.

## 3. Experimental Demonstration

The methodology was also tested using experimental BOS data of the flow induced by a nanosecond spark plasma discharge. A spark discharge of nanosecond duration leads to the rapid deposition of heat in the electrode gap leading to the development of a complex flow field with large thermal gradients. The experimental details corresponding to the dataset used in this assessment are reported in the work by Singh et. al. [15]. The BOS measurements were performed by imaging a random dot pattern (fabricated from sand-blasted aluminum) in the presence of flow induced by a spark across a 5 mm electrode gap. The dot pattern and flow were imaged at a magnification of 0.8 and a frame rate of 20 kHz with a 3.18 cm separation between the target and the electrodes. More details of the experimental setup can be found in Singh et. al. [15].

The dot pattern images were processed using the standard cross-correlation (SCC) method with multi-grid window deformation [16]. The window sizes were varied from 64 to 48 to 32 pixels over 4 passes with an overlap of 50% resulting in a final grid resolution of 16 pixels. The displacement uncertainties were calculated using the MC method described earlier, and the vector field is validated using Universal Outlier Detection (UOD) [17].

In the case of the experimental data, the extent of the path integration in Equation (1) is not known, as the flow is three-dimensional and only one view is presently available. Therefore, the displacements were used to calculate the *projected* density gradients, $\nabla \rho_p = \int \nabla \rho \, dz$, thereby constituting a 2D simplification of the 3D field. The gradients were then spatially integrated using the Poisson and WLS methods, with the weights assigned based on the projected density gradient uncertainty $\sigma_{\nabla \rho_p}$ for the latter, to yield the *projected* density field $\rho_p = \int \rho \, dz$. Dirichlet boundary conditions were imposed at the mid-points of the left and right boundaries and Neumann conditions were imposed elsewhere. Further, the Dirichlet density values were set to zero to calculate the 'relative' projected density field with respect to the ambient. During the experiment, the field of view was large enough (= 1 electrode gap on either side of the spark) to ensure that the left and right boundaries were far from the induced flow and truly in the ambient. While the analysis of Singh et. al. employed an Abel inversion procedure to further extract the radial density field, that was not performed here, and instead a direct comparison was performed on the projected density field to avoid introducing additional downstream steps/variables in the comparison.

The dot pattern displacements are shown in Figure 4(a) and (b) for two time instants, and the largest displacements occur at the boundary of the hot gas kernel as this corresponds to the largest temperature/density gradients. The kernel is initially cylindrical and deforms into a more complex shape at later times. The corresponding density field calculated using Poisson are shown in (c) and (d), and the density from WLS are shown in (e) and (f) for the same time instants. It is seen that the density is lower inside the gas kernel, corresponding to a higher gas temperature, as expected, and further that the two schemes result in similar density fields. However, it will be seen that there is a significant difference in the density uncertainty field.

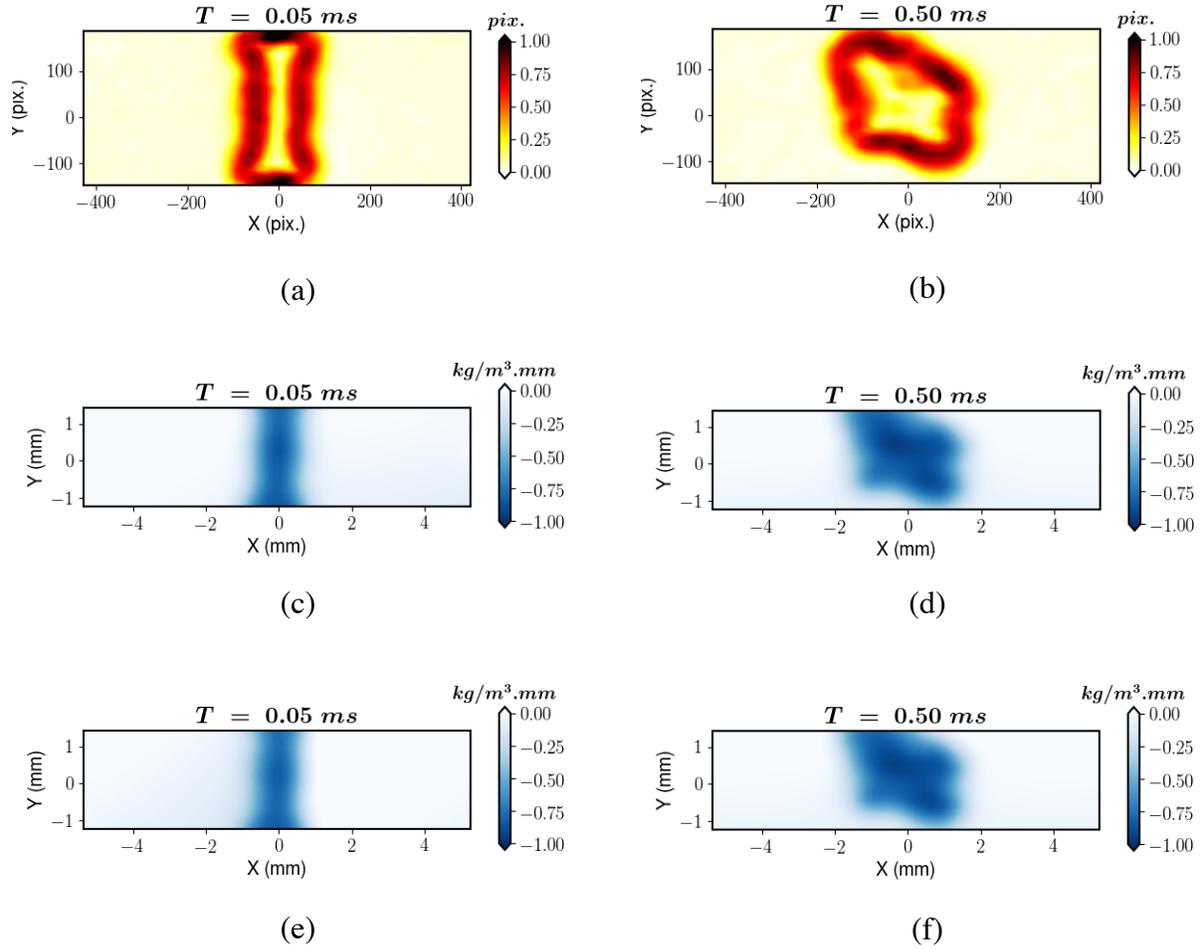

Figure 4. Flow induced by a spark discharge at two time instants. (a), (b): Instantaneous displacement fields and (c), (d): density fields obtained using Poisson, and (e), (f): density fields obtained using WLS. Plots (a), (c) and (e) correspond to the same time instant, as do (b), (d) and (f).

The corresponding uncertainties in the displacement fields are shown in Figure 5(a) and (b), and the displacement uncertainty is highest within the region occupied by the hot gas kernel, with the noise amplification ratio varying from 5–10 in this region. This is expected because in addition to higher displacements, the displacement gradients are also expected to be higher in this region, leading to a rise in the uncertainty. This uncertainty in the displacement is then propagated through the density integration process using the methodology described in [9] to result in the density uncertainty fields shown in Figure 5(c) – (f). Since the true density is not known for this experiment (as is the case with most experiments), the density *uncertainty* will be used in place of the density random error as the metric to compare the integration methods. Similar to the analysis with synthetic images, it is again seen that the density uncertainty from WLS is again lower and confined to the region within the hot gas kernel as opposed to that from Poisson where the uncertainty spreads to all regions in the domain.

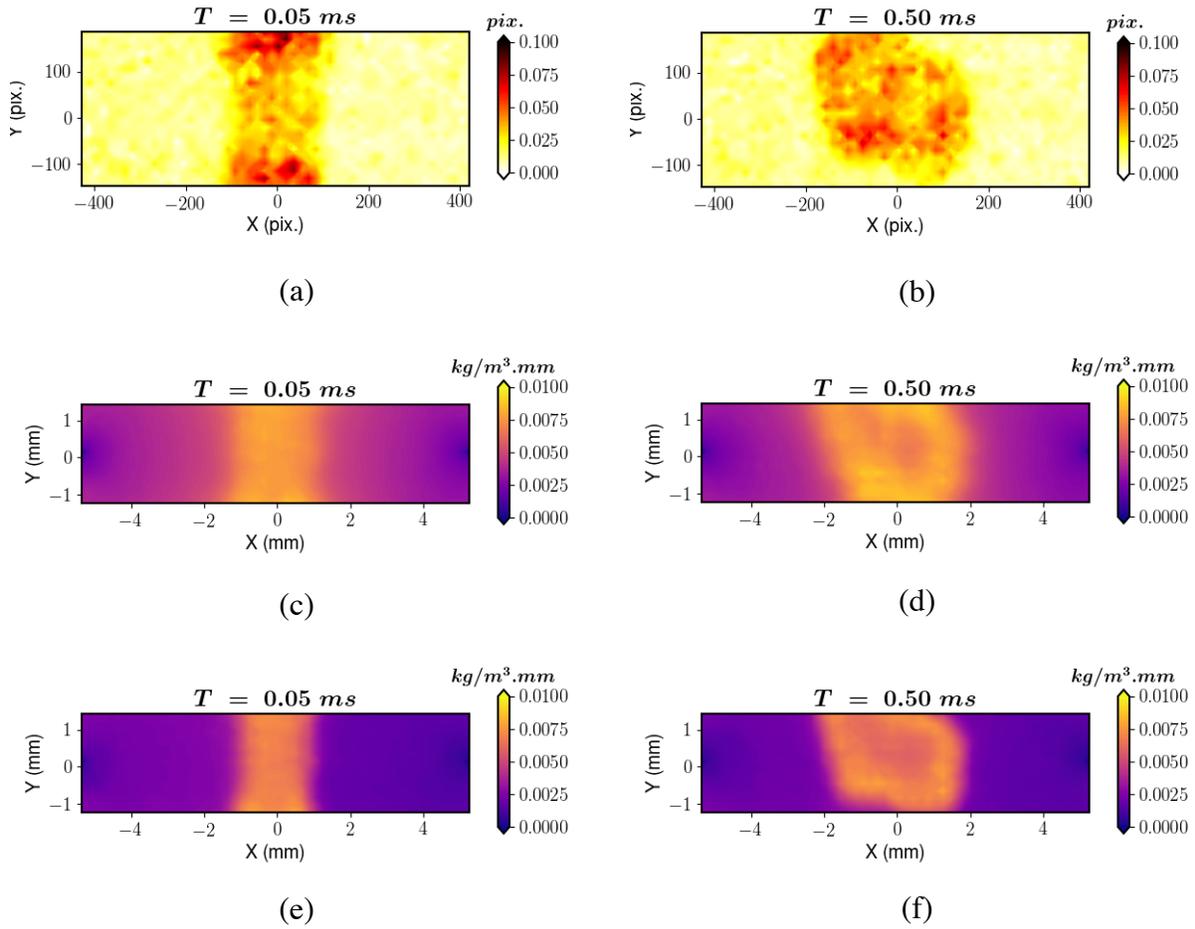

Figure 5. Instantaneous spatial distribution of the displacement uncertainty (a), (b) and the density uncertainty fields obtained from Poisson (c), (d) and WLS (e),(f) methods. Plots (a), (c) and (e) correspond to the same time instant, as do (b), (d), and (f).

In addition, the density uncertainty fields from 20 such snapshots of the flow were used to calculate the PDF and CDF distributions for the Poisson and WLS methods. The PDF with RMS of the uncertainty is shown in Figure 6(a) and the CDF with the 90$^{th}$ percentile is shown in in Figure 6 (b). It is seen that WLS reduces the RMS of the density uncertainty by 30% and the 90$^{th}$ percentile of the density by about 25%, thereby improving the overall precision in the density estimation.

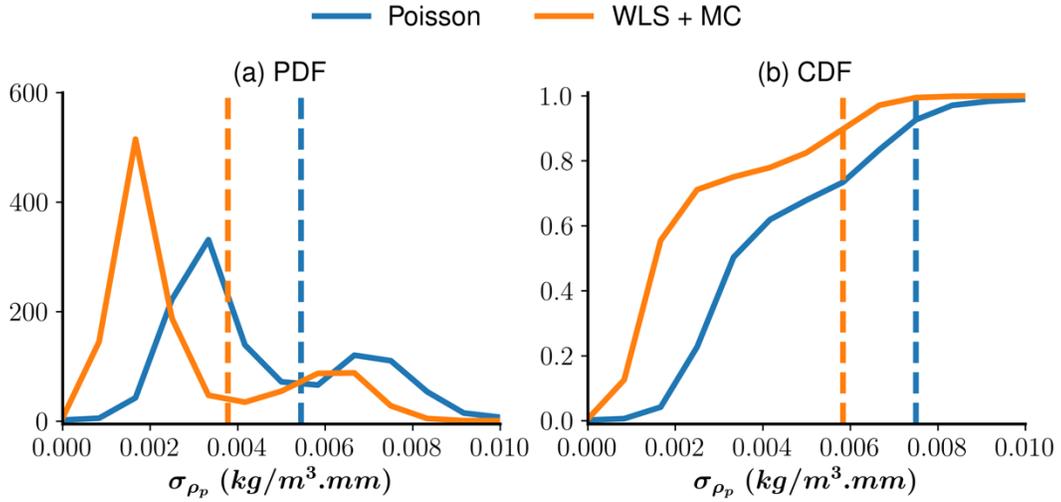

Figure 6. (a) PDF and (b) CDF of the density uncertainty associated with the two integration schemes for flow induced by the spark discharge.

## 4. Conclusion

In conclusion, we presented a weighted least squares (WLS) based density integration procedure in which weights are assigned to density gradient measurements based on the corresponding measurement uncertainty to improve the overall precision of the density integration procedure. Results from the synthetic image analysis showed that WLS was able to constrain the spread of the density random error compared to the Poisson solver and reduced the RMS error by 80% for the highest noise level. It was also seen that weights based on the Moment of Correlation uncertainty quantification scheme performed just as well as when weights were based on the displacement random error, thereby demonstrating that the displacement/density gradient uncertainty is a valid reference for assigning the weights. From the experimental images of flow induced by a spark plasma discharge, it was seen that WLS reduced the RMS uncertainty by 30% in comparison to Poisson, thus producing more precise density estimation.

Further improvement of the method can be achieved by accounting for the covariance in the displacement estimation procedure, to be used for assigning weights in a Generalized Least Squares (GLS) integration framework. Recent work in pressure integration has shown that GLS can further reduce the errors and uncertainties when the covariance information is available [8]. Work is ongoing to develop a methodology to estimate the covariance in displacement estimation between neighboring grid points from the cross-correlation plane for PIV/BOS measurements. Work is also ongoing on estimating the displacement uncertainty for tracking-based BOS measurements using the ratio of dot diameters in the reference and gradient images [10], and this is another avenue for the application of WLS based density integration.

## Acknowledgment

Bhavini Singh is acknowledged for help in conducting the spark discharge experiment. This material is based upon work supported by the U.S. Department of Energy, Office of Science, Office of Fusion Energy Sciences under Award Number DE-SC0018156.